\documentclass[letterpaper,12pt]{article}

\pdfoutput=1
\usepackage{graphicx}
\usepackage{amsmath}
\usepackage{amsfonts}
\usepackage{hyperref}

\makeatletter
\renewcommand\section{\@startsection {section}{1}{\z@}%
{-3.5ex \@plus -1ex \@minus -0.2ex}%
{2.3ex \@plus 0.2ex}%
{\normalfont\normalsize\bfseries}}

\renewcommand\subsection{\@startsection{subsection}{2}{\z@}%
{-3.25ex \@plus -1ex \@minus -0.2ex}%
{1.5ex \@plus 0.2ex}%
{\normalfont\normalsize\bfseries}}

\def\@seccntformat#1{\csname the#1\endcsname.\quad}
\makeatother

\begin{document}

\setlength{\baselineskip}{3.75ex}

\noindent
\textbf{\LARGE A revision to the theory of organic}\\[2ex]
\textbf{\LARGE fiducial inference}\\[3ex]

\noindent
\textbf{Russell J. Bowater}\\
\emph{Independent researcher, Sartre 47, Acatlima, Huajuapan de Le\'{o}n, Oaxaca, C.P.\ 69004,
Mexico. Email address: as given on arXiv.org. Twitter profile:
\href{https://twitter.com/naked_statist}{@naked\_statist}\\ Personal website:
\href{https://sites.google.com/site/bowaterfospage}{sites.google.com/site/bowaterfospage}}
\\[2ex]

\noindent
\textbf{\small Abstract:}
{\small A principle is modified that underlies the theory of organic fiducial inference as this
theory was presented in an earlier paper. This modification, which is arguably a natural one to
make, allows Bayesian inference to sometimes have a minor role within the theory in question and,
as a consequence, allows more information from the data to be incorporated into the way a full
conditional fiducial density is defined in certain cases. The new version of the principle
concerned is applied to examples that were analysed previously using the older version of this
principle.}
\\[3ex]

\vspace{3ex}
\section{Introduction}

The purpose of this paper is to present a modification to the theory of organic fiducial inference
as this theory was outlined in Bowater~(2019). The main change that will be put forward is one that
allows Bayesian inference to sometimes have a minor role within this theory. To avoid too much
repetition of the content of Bowater~(2019), only the parts of the theory in question that have
been altered will be discussed in detail. However, consideration will be given to the reader who
does not want to spend too much time referring back to this earlier paper.

In this regard, some of the concepts and principles that underlie the theory of organic fiducial
inference being referred to will be developed in the sections that immediately follow (i.e.\
Sections~\ref{sec2} to~\ref{sec3}) in a revised way to how they were developed in Bowater~(2019).
The version of organic fiducial inference put forward in these sections will then, in
Section~\ref{sec6}, be applied to examples that were originally analysed in Bowater~(2019) using
the version of this theory of inference that was outlined in this earlier paper.

\vspace{3ex}
\section{Theory of organic fiducial inference}
\label{sec1}

\vspace{1ex}
\subsection{Sampling model and data generation}
\label{sec2}

It will be assumed, in general, that the data set to be analysed $x=\{x_i: i=1,2,\ldots,n\}$ was
generated by a sampling model that depends on a set of unknown parameters $\theta = \{\theta_i:
i=1,2,\ldots,k\}$, where each $\theta_i$ is a one-dimensional variable.
Let the joint density or mass function of the data given the true values of the parameters $\theta$
be denoted as $g(x\,|\,\theta)$.
For the moment, though, we will assume that the only unknown parameter in the model is
$\theta_j$\hspace{0.05em}, either because there are no other parameters in the model, or because
the true values of the parameters in the set $\theta_{-j}= \{\theta_1, \ldots, \theta_{j-1},
\theta_{j+1}, \ldots, \theta_k \}$ are known.

To begin with, let us define more precisely the concept of a fiducial statistic in comparison to
how this type of statistic was defined in Bowater~(2019).

\vspace{3ex}
\noindent
\textbf{Definition 1: A fiducial statistic}

\vspace{1ex}
\noindent

Let $V(x)$ be a set of univariate statistics that together form a sufficient or approximately
sufficient set of statistics for the parameter $\theta_j$. To clarify, a set of statistics will be
defined as being an approximately sufficient set of statistics for $\theta_j$ if conditioning the
distribution of the data set of interest $x$ on this set of statistics leads to a distribution of
this data set $x$ that does not depend heavily on the value of the parameter $\theta_j$.

Now, if a given set $V(x)$, as just defined, only contains one statistic that is not an ancillary
statistic, then that statistic will be called the fiducial statistic $Q(x)$ of this set $V(x)$.
Assuming that the set $V(x)$ is of this nature, we will denote all the statistics in $V(x)$ except
for the statistic $Q(x)$ as the set of statistics $U(x) = \{U_i(x): i=1,2,\ldots,m\}$, and we will
refer to the statistics in this set as being the ancillary complements of $Q(x)$.

Note that often it may be possible to find a set $V(x)$ that consists of just a single sufficient
statistic for $\theta_j$\hspace{0.05em}, and of course in these cases, the fiducial statistic
$Q(x)$ will be this sufficient statistic and the set $U(x)$ will be empty. On other occasions, it
may be possible to find a single statistic that could sensibly be regarded as being an approximate
sufficient statistic for $\theta_j$\hspace{0.05em}, and if this is the case, we may decide to treat
this statistic as being the fiducial statistic $Q(x)$ for the example in question.
In this latter type of scenario, if may be appropriate to choose the set $U(x)$ to be empty or
non-empty. An example of a statistic that may often be regarded as being an approximately
sufficient statistic for $\theta_j$ is any one-to-one function of a unique maximum likelihood
estimator of $\theta_j$.
The use of a maximum likelihood estimator as a fiducial statistic was illustrated in Section~5.7 of
Bowater~(2018) and Section~3.5 of Bowater~(2020).

\vspace{3ex}
We will now make an assumption about how the data $x$ were generated. Essentially the same
assumption was also made in Bowater~(2019). Here, we simply present this assumption in a slightly
different way.

\vspace{3ex}
\noindent
\textbf{Assumption 1: Data generating algorithm}

\vspace{1ex}
\noindent
Independent of the way in which the data set $x$ was actually generated, it will be assumed that
this data set was generated by the following algorithm:

\vspace{2ex}
\noindent
1) Simulate the values $u = \{u_i: i=1,2,\ldots,m\}$ of the ancillary complements
$U(x) = \{U_i(x): i=1,2,\ldots,m\}$, if any exist, of a given fiducial statistic $Q(x)$.

\vspace{2ex}
\noindent
2) Generate a value $\gamma$ for a continuous scalar random variable $\Gamma$, which has a density
function $\pi_0(\gamma)$ that does not depend on the parameter $\theta_j$.

\vspace{2ex}
\noindent
3) Determine a value $q(x)$ for the fiducial statistic $Q(x)$ by setting $\Gamma$ equal to $\gamma$
and $Q(x)$ equal to $q(x)$ in the following expression for the statistic $Q(x)$, which should
effectively define a probability density or mass function for this statistic\hspace{0.05em}:
\begin{equation}
\label{equ1}
Q(x)=\varphi(\Gamma, \theta_j, u)
\end{equation}
where:\\
a) the function $\varphi(\Gamma, \theta_j, u)$ has as arguments only the variable $\Gamma$ and
constants such as the parameter $\theta_j$ and \pagebreak the statistics $u$.\\
b) the density or mass function of $Q(x)$ defined by this equation is equal to what it would have
been if $Q(x)$ had been determined on the basis of the data set $x$ conditional on the variables
$U(x)$, if any exist, being equal to the values $u$.

\vspace{2ex}
\noindent
4) Generate the data set $x$ from the sampling density or mass function
$g(x\,|\,\theta_1,\theta_2,\ldots,\theta_k)$ conditioned on the statistic $Q(x)$ being equal to its
already generated value $q(x)$ and the variables $U(x)$, if any exist, being equal to the
values $u$.

\vspace{3ex}
In the context of the above algorithm, the variable $\Gamma$ will be referred to as the primary
random variable (primary r.v.), which is the way that this term was used in Bowater~(2018, 2019,
2020). To clarify, if it is possible, which it is in many cases, to rewrite this
algorithm so that, after the data set $x$ is generated from the sampling density or mass function
$g(x\,|\,\theta)$ by using some black-box procedure, the value $\gamma$ of the variable $\Gamma$ is
generated by setting it equal to a deterministic function of the data $x$ and the parameter
$\theta_j$\hspace{0.05em}, then $\Gamma$ would not be the primary r.v.\ in the context of this
alternative algorithm.

\vspace{3ex}
\subsection{Types of fiducial argument and expression of pre-data knowledge}
\label{sec4}

Let us once again present the definition of the strong fiducial argument as it was given in
Bowater~(2019).

\vspace{3ex}
\noindent
\textbf{Definition 2(a): Strong or standard fiducial argument}

\vspace{1ex}
\noindent
This is the argument that the density function of the primary r.v.\ $\Gamma$ after the data have
been observed, i.e.\ the post-data density function of $\Gamma$, should be equal to the pre-data
density function of $\Gamma$, i.e.\ the density function $\pi_0(\gamma)$ as defined in step~2 of
the algorithm in Assumption~1.

\vspace{3ex}
The definitions of the moderate and weak fiducial arguments will not be presented here. These two
arguments will be assumed to have the same definitions as given in Bowater~(2019).

To assist the reader though, let us present, once more, the definition of the global pre-data (GPD)
function as it was given in this earlier paper.

\pagebreak
\noindent
\textbf{Definition 3: Global pre-data (GPD) function}

\vspace{1ex}
\noindent
The global pre-data (GPD) function $\omega_G(\theta_j)$ may be any given non-negative and upper
bounded function of the parameter $\theta_j$.
It is a function that only needs to be specified up to a proportionality constant, in the sense
that, if it is multiplied by a positive constant, then the value of the constant is redundant.

\vspace{3ex}
Even though in the modified version of the theory of organic fiducial inference being currently
outlined, the basic definition of the local pre-data (LPD) function is the same as the definition
of this function given in Bowater~(2019), its role in this theory of inference will now be quite
distinct from the role it was given in this earlier paper. It is appropriate therefore to give,
once again, the basic definition of a LPD function along with some general comments about this type
of function.

\vspace{3ex}
\noindent
\textbf{Definition 4: Local pre-data (LPD) function}

\vspace{1ex}
\noindent
The local pre-data (LPD) function $\omega_L(\theta_j)$ may be any given non-negative function of
the parameter $\theta_j$ that is locally integrable over the space of this parameter. Similar to a
GPD function, it only needs to be specified up to a proportionality constant.

The role of the LPD function is to facilitate the completion of the definition of the joint
post-data density function of the primary r.v.\ $\Gamma$ and the parameter $\theta_j$ in cases
where using either the strong or moderate fiducial argument alone is not sufficient to achieve
this. For this reason, the LPD function is in fact redundant in many situations.

We describe such a function as being `local' because it is only used in the inferential process
under the condition that $\Gamma$ equals a specific value, and with this condition in place, the
act of observing the data $x$ will usually imply that the parameter $\theta_j$ must lie in a
compact set that is contained in quite a small region of the real line.
It will be seen that because of this, even if the LPD function is not redundant, its influence on
the inferential process will be, in the main, relatively minor.

\vspace{3ex}
\subsection{Principles for defining univariate fiducial density functions}
\label{sec5}

As was the case in Bowater~(2019), the fiducial density function of the parameter $\theta_j$ given
the data $x$ and conditional on all other parameters $\theta_{-j}$ being known, i.e.\ the density
function $f(\theta_j\,|\,\theta_{-j},x)$, will be defined according to two mutually consistent
principles. The first of these principles is the same as Principle~1 for defining the density
function of $\theta_j$ in question that was outlined in Bowater~(2019). To avoid repetition of the
content of this earlier paper, this principle will not be presented again here.

On the other hand, the second principle for defining the density function
$f(\theta_j\,|\,\theta_{-j},x)$ that will be advocated in the current paper is distinct in an
important respect from Principle~2 for performing this task that was outlined in Bowater~(2019).
Therefore, even though this second principle is similar to Principle~2 given in Bowater~(2019), it
now will be presented and discussed in detail. To assist the reader, we will point out that this
principle differs from Principle~2 of Bowater~(2019) due to an important change that is made to the
definition of the conditional density function that is denoted in both the present paper and
Bowater~(2019) as $\omega_*(\theta_j\,|\,\gamma)$.

\vspace{3ex}
\noindent
\textbf{Principle 2 for defining a full conditional fiducial density}

\vspace{1ex}
\noindent
To be able to use this principle, the following two conditions must be satisfied.

\vspace{2ex}
\noindent
\textbf{Condition 2(a)}

\vspace{1ex}
\noindent
Let $G_x$ and $H_x$ be, respectively, the sets of all the values of the primary r.v.\ $\Gamma$ and
the parameter $\theta_j$ for which the density functions of these variables must necessarily be
positive in light of having observed the data $x$.
To clarify, any set of values of $\Gamma$ or any set of values of $\theta_j$ that are regarded as
being impossible after the data have been observed can not be contained in the set $G_x$ or the set
$H_x$ respectively.

Given this notation, the present condition will be satisfied if
\begin{equation}
\label{equ2}
H_x = \{\hspace{0.1em} \theta_j : (\exists\hspace{0.1em} \gamma \in G_x)
[\hspace{0.05em}\theta_j\hspace{-0.05em} \in \theta_j(\gamma)\hspace{0.05em}]\hspace{0.1em} \}
\end{equation}
where $\theta_j(\gamma)$ is the set of values of the parameter $\theta_j$ that map on to the value
$\gamma$ for the variable $\Gamma$ according to equation~(\ref{equ1}) if the variable $Q(x)$ in
this equation is substituted by its observed value $q(x)$, and the values $u$, if there are any,
are held fixed at their observed values.
(To clarify, the predicate in the definition of the set on the right-hand side of
equation~(\ref{equ2}) means `there exists a $\gamma \in G_x$ such that $\theta_j \in
\theta_j(\gamma)$').

\pagebreak
\noindent
\textbf{Condition 2(b)}

\vspace{1ex}
\noindent
The GPD function $\omega_G(\theta_j)$ must be such that:
\[
\omega_G(\theta_j) = a\ \ \ \mbox{for all $\theta_j \in H_x$}
\]
where $a>0$.

\vspace{2ex}
Under Conditions~2(a) and~2(b), the full conditional fiducial density
$f(\theta_j\,|\,\theta_{-j},x)$ is defined by:
\vspace{1ex}
\begin{equation}
\label{equ3}
f(\theta_j\,|\,\theta_{-j},x) = \int_{\mbox{\footnotesize $\gamma\hspace{-0.15em}
\in\hspace{-0.15em} G_x$}}\hspace{0.1em} \omega_*(\theta_j\,|\,\gamma)\hspace{0.05em}
\pi_1 (\gamma)\hspace{0.05em} d\gamma
\vspace{1ex}
\end{equation}
where:\\
i) the post-data density $\pi_1(\gamma)$ is given by:
\vspace{0.5ex}
\begin{equation}
\label{equ4}
\pi_1(\gamma) = \left\{
\begin{array}{ll}
{\tt C}_0\hspace{0.1em} \pi_0(\gamma)\, & \mbox{if $\gamma \in G_x$}\\[1ex]
0 & \mbox{otherwise}
\end{array}
\right.
\vspace{0.5ex}
\end{equation}
in which ${\tt C}_0$ is a normalising constant and the density $\pi_0(\gamma)$ is as defined in
step~2 of the algorithm in Assumption~1.\\
ii) the conditional density function $\omega_*(\theta_j\,|\,\gamma)$ is given by:
\vspace{1ex}
\begin{equation}
\label{equ5}
\omega_*(\theta_j\,|\,\gamma) = \left\{
\begin{array}{ll}
{\tt C}_1\hspace{-0.03em}(\gamma)\hspace{0.1em}l(\theta_j\,|\,x)\hspace{0.1em}
\omega_L(\theta_j)\, &
\mbox{if $\theta_j \in \theta_j(\gamma)$}\\[1.25ex]
0 & \mbox{otherwise}
\end{array}
\right.
\vspace{1ex}
\end{equation}
in which $\omega_L(\theta_j)$ is the LPD function of $\theta_j$ as introduced by Definition~4, the
function $l(\theta_j\,|\,x)$ is the likelihood function of $\theta_j$ given the data $x$, the set
$\theta_j(\gamma)$ is as defined in Condition~2(a), and ${\tt C}_1\hspace{-0.03em}(\gamma)$ is a
normalising constant, which clearly must depend on the value of $\gamma$.

\vspace{3ex}
It can be seen that the density function $f(\theta_j\,|\,\theta_{-j},x)$ as defined by
equation~(\ref{equ3}) is formed by marginalising, with respect to $\gamma$, a joint density of the
primary r.v.\ $\Gamma$ and the parameter $\theta_j$ that is based on
$\omega_*(\theta_j\,|\,\gamma)$ being the conditional density of $\theta_j$ given $\gamma$, and on
$\pi_1 (\gamma)$ being the marginal density of $\Gamma$. Also, it is clearly the case that if
\begin{equation}
\label{equ6}
G_x = \{ \gamma: \pi_0(\gamma) > 0 \}
\pagebreak
\end{equation}
then the post-data density $\pi_1(\gamma)$ will be equal to the pre-data density $\pi_0(\gamma)$,
i.e.\ the fiducial density $f(\theta_j\,|\,\theta_{-j},x)$ is determined on the basis of the strong
fiducial argument, otherwise this fiducial density of $\theta_j$ is determined on the basis of the
moderate fiducial argument.
To clarify, in contrast to what was the case under Principle~1 for defining the density
$f(\theta_j\,|\,\theta_{-j},x)$ outlined in Bowater~(2019), the weak fiducial argument is never
used to make inferences about $\theta_j$.

Furthermore, we can observe that the density function $\omega_*(\theta_j\,|\,\gamma)$ defined in
equation~(\ref{equ5}) is the posterior density of $\theta_j$ when $\theta_j$ is conditioned to lie
in the set $\theta_j(\gamma)$ and when the height of the prior density of $\theta_j$ is
proportional to the height of the LPD function of $\theta_j$ over the set $\theta_j(\gamma)$.
The role of the LPD function of $\theta_j$ in constructing the fiducial density
$f(\theta_j\,|\,\theta_{-j},x)$ is therefore to specify how $\theta_j$ was distributed a priori
over those values of $\theta_j$ that are consistent with any given value of $\Gamma$.
For this reason, it is assumed that this LPD function is chosen to reflect what we believed about
the parameter $\theta_j$ before the data were observed.
As eluded to in Definition~4, the sets $\theta_j(\gamma)$ will usually be compact sets that are
wholly contained within quite small regions of the real line.

To clarify, the difference between the current definition of the density
$\omega_*(\theta_j\,|\,\gamma)$ and the earlier definition of this density function given in
Bowater~(2019) is that, in effect, this density function is now the posterior rather than the prior
density of $\theta_j$ when $\theta_j$ is conditioned to lie in the set $\theta_j(\gamma)$.
This modification therefore allows, in a certain sense, more information from the data to be
incorporated into the way the fiducial density $f(\theta_j\,|\,\theta_{-j},x)$ is constructed.
The issue as to whether this extra information is processed adequately by the method of inference
in question will be discussed in the next section.

Finally, it can be appreciated that, if Condition~2(b) is satisfied, then Principle~1 of
Bowater~(2019) is essentially a special case of the version of Principle~2 that has just been
presented.
In particular, we are able to see that if the necessary condition to use Principle~1 is satisfied,
i.e.\ Condition~1 of Bowater~(2019), then Condition~2(a) will be satisfied, and so if
Condition~2(b) also holds, then both conditions required to use Principle~2 will hold.
Also, under Condition~1 of Bowater~(2019), the density function $\omega_*(\theta_j\,|\,\gamma)$
could be regarded as converting itself into a point mass function at the value $\theta_j(\gamma)$,
and as a result, the joint density function of $\Gamma$ and $\theta_j$ in equation~(\ref{equ3})
effectively becomes a univariate density function.
Therefore, the integration of this latter function with respect to $\gamma$ in this equation would
be, under Condition~1 of Bowater~(2019), naturally regarded as being redundant, and so
equation~(\ref{equ3}) would, in effect, define the fiducial density $f(\theta_j\,|\,\theta_{-j},x)$
according to Principle~1 of the earlier paper in question.

\vspace{3ex}
\noindent
\textbf{Can fiducial and Bayesian inference work together?}

\vspace{1ex}
By using Principle~2 just outlined to determine the fiducial density
$f(\theta_j\,|\, \theta_{-j},x)$, it can be seen that we effectively need to combine fiducial
inference with Bayesian inference, and therefore, we may ask whether such a combination of two
methods of inference that are quite distinct is acceptable.
In this regard, it should first be pointed out that these two methods of inference do not directly
interfere with each other since the marginal post-data distribution of $\Gamma$ is determined by
solely using fiducial inference, while the post-data distribution of $\theta_j$ conditional on
$\gamma$ is determined by solely using Bayesian inference.
Nevertheless, the strong fiducial argument would be naturally invoked when very little or nothing
was known about the parameter of interest $\theta_j$ before the data were observed, which could be
taken as meaning that our amount of pre-data knowledge about $\theta_j$ should be below a level
that would allow this knowledge to be adequately represented by placing a probability distribution
over $\theta_j$.
Therefore, if the strong fiducial argument needs to be invoked in applying Principle~2 to determine
the density $f(\theta_j\,|\, \theta_{-j},x)$, this may stand in conflict with the need to express
pre-data knowledge about $\theta_j$ in the form of a LPD function over $\theta_j$.

On the other hand, in doing the best we can to apply the Bayesian method to any given problem of
inference in which there was very little pre-data knowledge about one of the parameters of interest
$\theta_j$\hspace{0.05em}, it is natural that the inferential procedure should take into account
the variation in the posterior distribution of all the model parameters $\theta$ over a wide range
of prior distributions for the parameter $\theta_j$\hspace{0.05em}, each of which may be considered
to loosely represent our pre-data knowledge about this parameter. We should note, of course, that
if we actually knew truly nothing about the parameter $\theta_j$ before the data were observed,
then, since we would arguably need to perform such a sensitivity analysis over all possible prior
distributions for this parameter, our post-data or posterior inferences about the parameter
$\theta_j$ would, in general, be completely uninformative.
However, this would not be the case if, in the context of using Principle~2, we performed such a
complete sensitivity analysis of the post-data density $f(\theta_j\,|\,\theta_{-j},x)$ with respect
to the LPD function of $\theta_j$ as of course this post-data density of $\theta_j$ is, in general,
mainly determined by fiducial inference and not Bayesian inference.
Moreover if, in the same context, we had possessed just a very small amount of pre-data knowledge
about $\theta_j$\hspace{0.05em}, then we could use a more constrained version of the type of
sensitivity analysis under discussion to take advantage of this knowledge, while at the same time
we may well still feel that this knowledge is not important enough for the use of the strong
fiducial argument to be made inappropriate.

In summary, a conflict between fiducial inference and Bayesian inference can be avoided when
applying Principle~2 by bearing in mind that the kind of pre-data knowledge we would need to have
about $\theta_j$ to make it sensible to invoke either the moderate or strong fiducial argument when
using this principle would be such that the variation in the fiducial density
$f(\theta_j\,|\,\theta_{-j},x)$ should be taken into account over a range of different LPD
functions of $\theta_j$\hspace{0.05em}, each of which may be considered as loosely representing our
pre-data knowledge about $\theta_j$. This would seem to make it awkward to use Principle~2 to
determine the fiducial density $f(\theta_j\,|\,\theta_{-j},x)$, however it will be very often the
case that the variation in this fiducial density of $\theta_j$ that is seen in this kind of
sensitivity analysis will be very minor or negligible.

\vspace{3ex}
\subsection{Extending the theory to the general case}
\label{sec3}

\vspace{1ex}
It will be assumed that the theory of organic fiducial inference is extended to the case in which
all the parameters $\theta = \{ \theta_1, \theta_2, \ldots, \theta_k\}$ in the sampling model are
unknown using exactly the same type of approach as described in Bowater~(2019), which is
essentially the approach that was originally put forward in Bowater~(2018). In other words, we
first construct each of the fiducial densities in the complete set of full conditional fiducial
densities for the parameters $\theta$, i.e.\ the set of fiducial densities:
\[
f(\theta_j\,|\,\theta_{-j},x)\ \ \ \mbox{for $j=1,2,\ldots,k$}
\]
by applying Principle~1 for this type of task (as outlined in Bowater~2019) or Principle~2 (as has
just been outlined), or any related principle (such as the ones outlined in Sections~7.2 and~8 of
Bowater~2019), and then we determine the joint fiducial density of all the parameters $\theta$
based on these full conditional densities using either an analytical method or an approach based on
the Gibbs sampling algorithm that was outlined in Bowater~(2018, 2019).

For further details on the extension of organic fiducial inference to the general case under
discussion, the reader is therefore referred to the two papers that have just been highlighted.

\vspace{3ex}
\section{Examples}
\label{sec6}

In the two sections that follow, we will show how Principle~2 as outlined in Section~\ref{sec5} can
be used to determine the fiducial density of a single parameter of interest in two examples. We
will consider first the case where the sampling distribution is binomial, and second the case where
this distribution is Poisson. These two examples were examined in Section~6.1 and~6.2 of
Bowater~(2019) using the older version of Principle~2 that was advocated in that paper. Therefore,
it may be helpful to compare the analyses that are about to be presented of these examples with the
analyses of the same examples presented in this earlier paper.

\vspace{2ex}
\subsection{Inference about a binomial proportion}
\label{sec7}

As just mentioned, we will begin by assuming that the sampling distribution is binomial. More
precisely, let us consider the problem of making inferences about the population proportion of
successes $p$ on the basis of observing $x$ successes in $n$ trials, where the probability of
observing any given number of successes $y$ is specified by the binomial mass function in this
case, i.e.\ the function:
\vspace{0.5ex}
\[
g_0(y\,|\,p) = \binom{n}{y} p^{\hspace{0.05em}y} (1-p)^{n-y}\ \ \ \mbox{for $y=0,1,\ldots,n$}
\vspace{0.5ex}
\]
We will suppose that nothing or very little was known about the proportion $p$ before the data were
observed.

As clearly the value $x$ is a sufficient statistic for the proportion $p$, it can therefore be
assumed to be the fiducial statistic $Q(x)$ in this example.
Based on this assumption, equation~(\ref{equ1}) can be expressed as:
\vspace{0.5ex}
\begin{equation}
\label{equ7}
x=\varphi(\Gamma,p)= \min \left\{ z: \Gamma < \mbox{\large $\sum$}_{\mbox{\footnotesize
$y\hspace{-0.25em}=\hspace{-0.25em}0$}}^{\mbox{\footnotesize $z$}}\hspace{0.3em}g_0(y\,|\,p)
\right\}
\vspace{0.5ex}
\end{equation}
where the primary r.v.\ $\Gamma$ has a uniform distribution over the interval $(0,1)$.
Under the assumption of there having been no or very little pre-data knowledge about $p$, it is
quite natural that the GPD function has the following form: $\omega_G(p)=a$\hspace{0.1em} if
$0 \leq p \leq 1$ and zero otherwise, where $a>0$.
Let us point out that for whatever choice is made for the GPD function of $p$ and whatever turns
out to be the value of $x$, we will never be able to apply Principle~1 of Bowater~(2019) to
determine the fiducial density of $p$ (see Bowater 2019 for more details).
On the other hand, equation~(\ref{equ7}) together with the GPD function for $p$ just specified will
satisfy Condition~2(a) of Section~\ref{sec5} for all possible values of $x$, and since
Condition~2(b) will also hold for all $x$, Principle~2 of this earlier section can always be
applied to the specific case of current interest.
Furthermore, as the condition in equation~(\ref{equ6}) will also be satisfied, inferences will be
made about the proportion $p$ under this principle by using the strong fiducial argument.

In particular, by placing the present case in the context of the general definition of the fiducial
density $f(\theta_j\,|\,\theta_{-j},x)$ given in equations~(\ref{equ3}), (\ref{equ4})
and~(\ref{equ5}), we obtain the following expression for the fiducial density of
$p$\hspace{0.1em}:
\vspace{0.5ex}
\begin{equation}
\label{equ8}
f(p\,|\, x) = \int_0^1 \omega_*(p\,|\,\gamma) \pi_1(\gamma) d\gamma =
\int_0^1 \omega_*(p\,|\,\gamma) d\gamma
\end{equation}
where
\vspace{1ex}
\begin{equation}
\label{equ9}
\omega_*(p\,|\,\gamma) = \left\{
\begin{array}{ll}
{\tt C}_1\hspace{-0.03em}(\gamma)\hspace{0.1em}p^{\hspace{0.05em}x} (1-p)^{n-x} \omega_L(p)\, &
\mbox{if $p \in p(\gamma)$}\\[1ex]
0 & \mbox{otherwise}
\end{array}
\right.
\vspace{1.5ex}
\end{equation}
in which $p(\gamma)$ is the set of values of $p$ that map on to the value $\gamma$ for the primary
r.v.\ $\Gamma$ according to equation~(\ref{equ7}) given the observed value of $x$.
Of course, to be able to complete this definition of the fiducial density $f(p\,|\, x)$, a LPD
function for $p$, i.e.\ the function $\omega_L(p)$, needs to be specified.
Observe that any choice for this function that satisfies the loose requirements of Definition~4 and
is positive for all values of $p$ will lead to a fiducial density $f(p\,|\,x)$ that is valid for
any $n \geq 1$ and any $x=0,1,\ldots, n$. Nevertheless, to facilitate the comparison of the
current analysis of the example under discussion with the analysis of the same example given in
Bowater~(2019), we will choose to highlight the same two LPD functions of $p$ that were highlighted
in this earlier paper, namely the LPD functions of $p$ that are \pagebreak defined by:
\begin{equation}
\label{equ10}
\omega_L(p) = b\ \ \ \mbox{if $0 \leq p \leq 1$ and zero otherwise}
\end{equation}
where $b>0$, and by:
\begin{equation}
\label{equ11}
\omega_L(p) = 1 / \sqrt{p(1-p)}\ \ \ \mbox{if $0 \leq p \leq 1$ and zero otherwise}
\end{equation}

Finally, we should point out that, although directly computing the density function $f(p\,|\,x)$ is
quite complicated, random values can, in general, be easily generated from this density function
using the same method of sampling from a fiducial density of this type that was relied upon in
Bowater~(2019). In particular, to obtain one random value from this density function of $p$, we
only need to generate a value $\gamma$ for the primary r.v.\ $\Gamma$ from its post-data density
function, i.e.\ a uniform density over the interval $(0,1)$, and then draw a value for the
proportion $p$ from the conditional density $\omega_*(p\,|\,\gamma)$.

To give an illustration of the results that are obtained by generating values from the fiducial
density $f(p\,|\,x)$ defined by equations~(\ref{equ8}) and~(\ref{equ9}), the histograms in
Figures~1(a) and~1(b) were each formed on the basis of one million independent random values drawn
from this fiducial density of $p$, with $n$ being equal to 10 and the observed $x$ being equal to
one. These values for $n$ and $x$ are identical to the values that were assigned to these
quantities in generating the results that are illustrated by the histograms in Figures 1(a)
and~1(b) of Bowater~(2019). Similar to this earlier paper, the results conveyed by the histogram in
Figure~1(a) of the present paper depend on choosing the LPD function of $p$ to be the one given in
equation~(\ref{equ10}), while the results in Figure~1(b) depend on this function being as defined
in equation~(\ref{equ11}).

\begin{figure}[t]
\begin{center}
\makebox[\textwidth]{\includegraphics[width=7in]{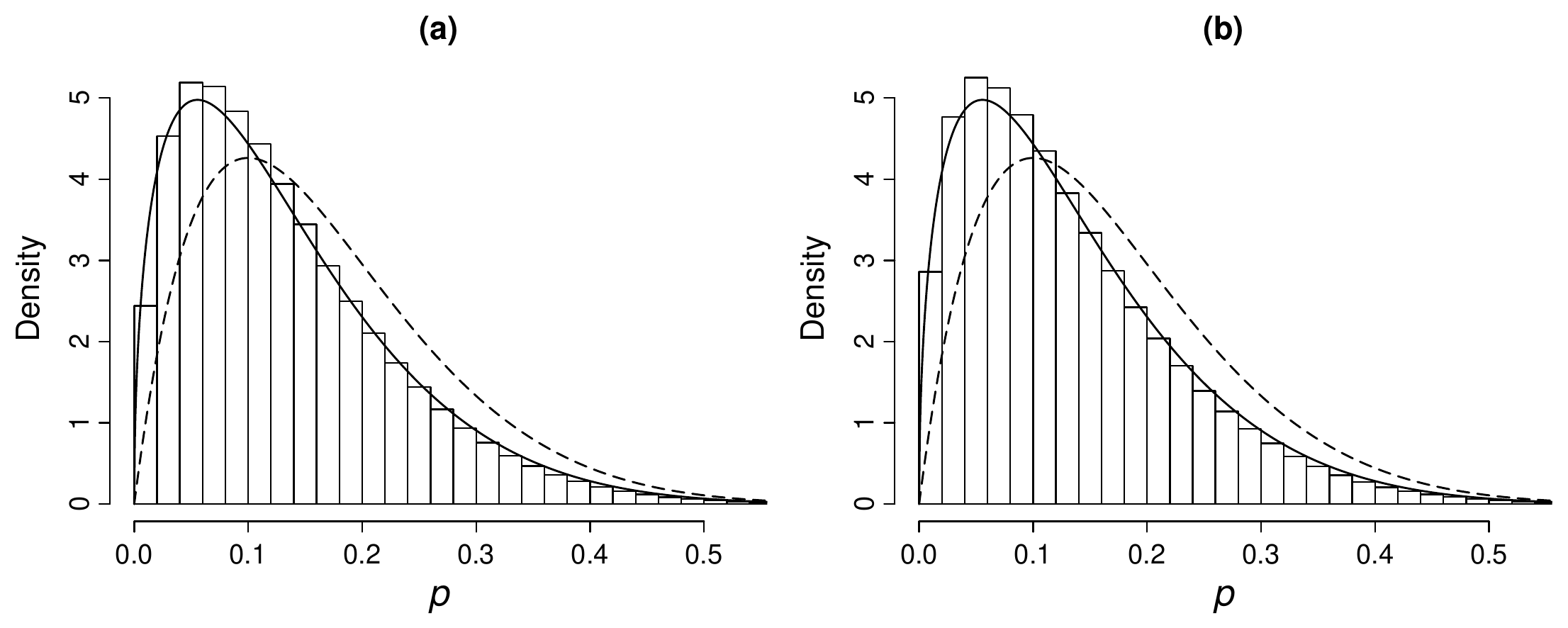}}
\caption{{\small Histograms representing samples from fiducial densities of a binomial proportion
when $n=10$ and $x=1$}}
\end{center}
\end{figure}

On the basis of the same data, the dashed curves in these figures represent the posterior density
for $p$ that (under the Bayesian paradigm) corresponds to the prior density for $p$ being a uniform
density on the interval $(0,1)$, while the solid curves in these figures represent the posterior
density for $p$ that corresponds to the prior density for $p$ being the Jeffreys prior for the case
in question, i.e.\ the prior density for $p$ that is proportional to the function of $p$ in
equation~(\ref{equ11}).

It can be seen from these figures that, although the posterior density for $p$ is highly sensitive
to which of the two prior densities for $p$ is used, the fiducial density of $p$ barely moves
depending on whether the LPD function of $p$ is proportional to the uniform prior density being
referred to, or whether it is proportional to the Jeffreys prior density for this case.
Furthermore, we can observe that the two fiducial densities for $p$ being considered are both
loosely approximated by the posterior density for $p$ that is based on the Jeffreys prior density
in question, except for values of $p$ that are close to the modes of these two fiducial densities.

To give an additional example, the histograms in Figures~2(a) and~2(b) were again each formed on
the basis of one million independent random values drawn from the fiducial density $f(p\,|\,x)$,
but this time the number of trials $n$ was set equal to 20 and the number of successes $x$ was set
equal to $2$, meaning of course that, as in the previous example, the sample proportion was equal
to 0.1. Once more, on the basis of the same data, the dashed curves in these figures represent the
posterior density for $p$ that corresponds to the prior density for $p$ being a uniform density on
the interval $(0,1)$, while the solid curves in these figures represent the posterior density for
$p$ that corresponds to the prior density for $p$ being the Jeffreys prior for the case in
question. The same type of comments can be made about these two figures as were made about
Figures~1(a) and~1(b) except that now we can see that the two fiducial densities for $p$ in these
figures are closely rather than loosely approximated by the posterior density for $p$ that is based
on the Jeffreys prior density for this case.

\begin{figure}[t]
\begin{center}
\makebox[\textwidth]{\includegraphics[width=7in]{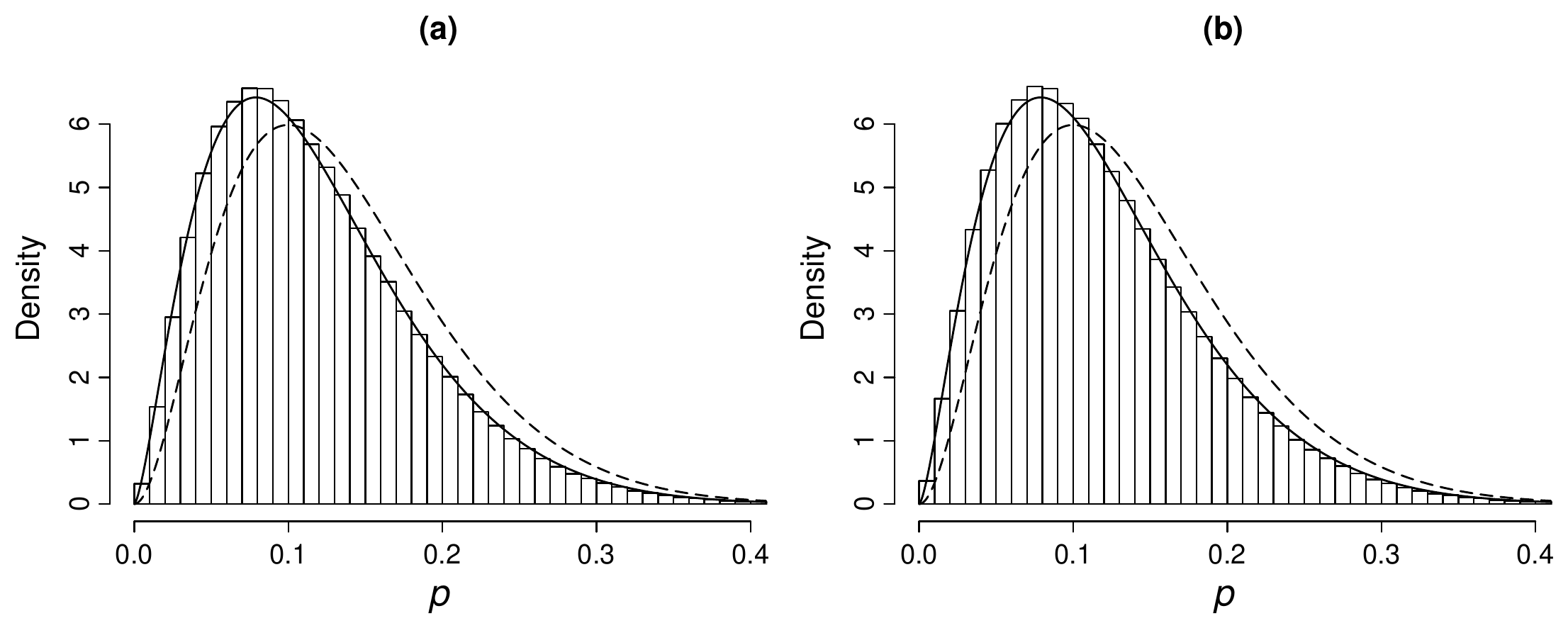}}
\caption{{\small Histograms representing samples from fiducial densities of a binomial proportion
when $n=20$ and $x=2$}}
\end{center}
\end{figure}

On the basis of further simulation results that are not reported here, it would be reasonable to
make the conjecture that, for any given sample proportion $x/n$, the fiducial density $f(p\,|\,x)$
will approximate more and more closely the posterior density for $p$ that is based on the Jeffreys
prior density for this case as the number of trials $n$ gets gradually larger, and that this
tendency is not simply due to the effect of the central limit theorem. However, no formal proof of
a result of this nature will be given here.

Observe that although the LPD functions in equations~(\ref{equ10}) and~(\ref{equ11}) are quite
different, the lack of a substantial amount of pre-data knowledge about the proportion $p$ may
often mean that we should take into account the variation in the fiducial density $f(p\,|\,x)$ over
a much wider range of LPD functions of $p$ before any final conclusions about the precise nature of
the post-data distribution of $p$ are made. This observation is completely in accordance with the
general recommendations concerning the use of Principle~2 for determining any given fiducial
density $f(\theta_j\,|\,\theta_{-j},x)$ that were outlined in Section~\ref{sec5}.

\vspace{3ex}
\subsection{Inference about a Poisson event rate}

We now will assume that the sampling distribution is Poisson. More precisely, let us consider the
problem of making inferences about an unknown event rate $\tau$ on the basis of observing $x$
events over a time period of length $t$, where the probability of observing any given number of
events $y$ over a period of this length is specified by a function that has the form of a Poisson
mass function, in particular the following function:
\[
g_1(y\,|\,\tau) = (\tau^{\hspace{0.05em}y} / y!)  \exp(-\tau)\ \ \ \mbox{for $y=0,1,2,\ldots$}
\]

Again, since the data set to be analysed consists of a single value $x$, this value can be assumed
to be the fiducial statistic $Q(x)$ in this example. Based on this assumption, we can express
equation~(\ref{equ1}) in a way that is similar to how this formula was expressed in
equation~(\ref{equ7}), in particular in the following way:
\vspace{0.5ex}
\begin{equation}
\label{equ12}
x=\varphi(\Gamma,\tau)= \min \left\{ z: \Gamma < \mbox{\large $\sum$}_{\mbox{\footnotesize
$y\hspace{-0.25em}=\hspace{-0.25em}0$}}^{\mbox{\footnotesize $z$}}\hspace{0.3em}g_1(y\,|\,\tau)
\right\}
\vspace{0.5ex}
\end{equation}
where again the primary r.v.\ $\Gamma$ has a uniform distribution over the interval $(0,1)$.

As it will be once more assumed that there was no or very little pre-data knowledge about the
parameter of interest, i.e.\ the event rate $\tau$ in this case, the GPD function will again be
specified in the following way: $\omega_G(\tau)=a$ if $\tau>0$ and zero otherwise, where $a>0$.
Similar also to the previous problem, it can be easily appreciated that the nature of
equation~(\ref{equ12}) implies that Principle~1 of Bowater~(2019) can never be applied to
determine the fiducial density of $\tau$ for any choice of the GPD function of $\tau$. However, the
specific choice that has been made for this latter function means that again Principle~2 of
Section~\ref{sec5} can be applied to the case at hand for all possible values of $x$, and
furthermore, inferences will be made about $\tau$ under this principle by using the strong fiducial
argument.

In particular, by placing the present case in the context of the general definition of the fiducial
density $f(\theta_j\,|\,\theta_{-j},x)$ given in equations~(\ref{equ3}), (\ref{equ4})
and~(\ref{equ5}), we obtain the following expression for the fiducial density of
$\tau$\hspace{0.1em}:
\vspace{0.5ex}
\[
f(\tau\,|\, x) = \int_0^1 \omega_*(\tau\,|\,\gamma) \pi_1(\gamma) d\gamma =
\int_0^1 \omega_*(\tau\,|\,\gamma) d\gamma
\]
where
\vspace{1ex}
\[
\omega_*(\tau\,|\,\gamma) = \left\{
\begin{array}{ll}
{\tt C}_1\hspace{-0.03em}(\gamma)\hspace{0.1em}\tau^{\hspace{0.05em}x}\hspace{-0.05em}
\exp(-\tau)\hspace{0.05em}\omega_L(\tau)\, & \mbox{if $\tau \in \tau(\gamma)$}\\[1ex]
0 & \mbox{otherwise}
\end{array}
\right.
\vspace{1ex}
\]
in which $\tau(\gamma)$ is the set of values of $\tau$ that map on to the value $\gamma$ for the
primary r.v.\ $\Gamma$ according to equation~(\ref{equ12}) given the observed value of $x$.
Of course, similar to the previous problem, a LPD function $\omega_L(\tau)$ is required so that the
definition of the fiducial density $f(\tau\,|\,x)$ can be completed.
Although any choice for this LPD function that conforms to Definition~4 and is positive for all
values of $\tau$ will imply that the fiducial density in question is valid for any
$x=0,1,2,\ldots$, let us choose to highlight the consequences of using the two LPD functions for
$\tau$ that are defined by:
\begin{equation}
\label{equ13}
\omega_L(\tau) = b\ \ \ \mbox{if $\tau>0$ and zero otherwise}
\end{equation}
where $b>0$, and by:
\begin{equation}
\label{equ14}
\omega_L(\tau) = 1 / \sqrt{\tau}\ \ \ \mbox{if $\tau>0$ and zero otherwise}
\vspace{1ex}
\end{equation}
These two LPD functions of $\tau$ are indeed the same as the two LPD functions of $\tau$ that were
highlighted when the current example was analysed in Bowater~(2019).

In relation to the issue being discussed, Figures~3(a) and~3(b) each show a histogram that was
formed on the basis of one million independent random values drawn from the fiducial density
$f(\tau\,|\,x)$ using the same simple simulation method as outlined in Section~\ref{sec7}, with the
observed count $x$ assumed to be equal to 2, and with the LPD functions of $\tau$ that underlie the
results conveyed by the histograms in these two figures being defined by equations~(\ref{equ13})
and~(\ref{equ14}) respectively.
On the basis also of $x$ being equal to 2, the dashed curves in these figures represent the
posterior density for $\tau$ that corresponds to the prior density for $\tau$ being the function of
$\tau$ given in equation~(\ref{equ13}), while the solid curves in these figures represent this
posterior density when the prior density for $\tau$ is the function of $\tau$ given in
equation~(\ref{equ14}), i.e.\ the Jeffreys prior for the case in question.
It should be pointed out that the use of these two prior densities is controversial as they are
both improper.

\begin{figure}[t]
\begin{center}
\makebox[\textwidth]{\includegraphics[width=7in]{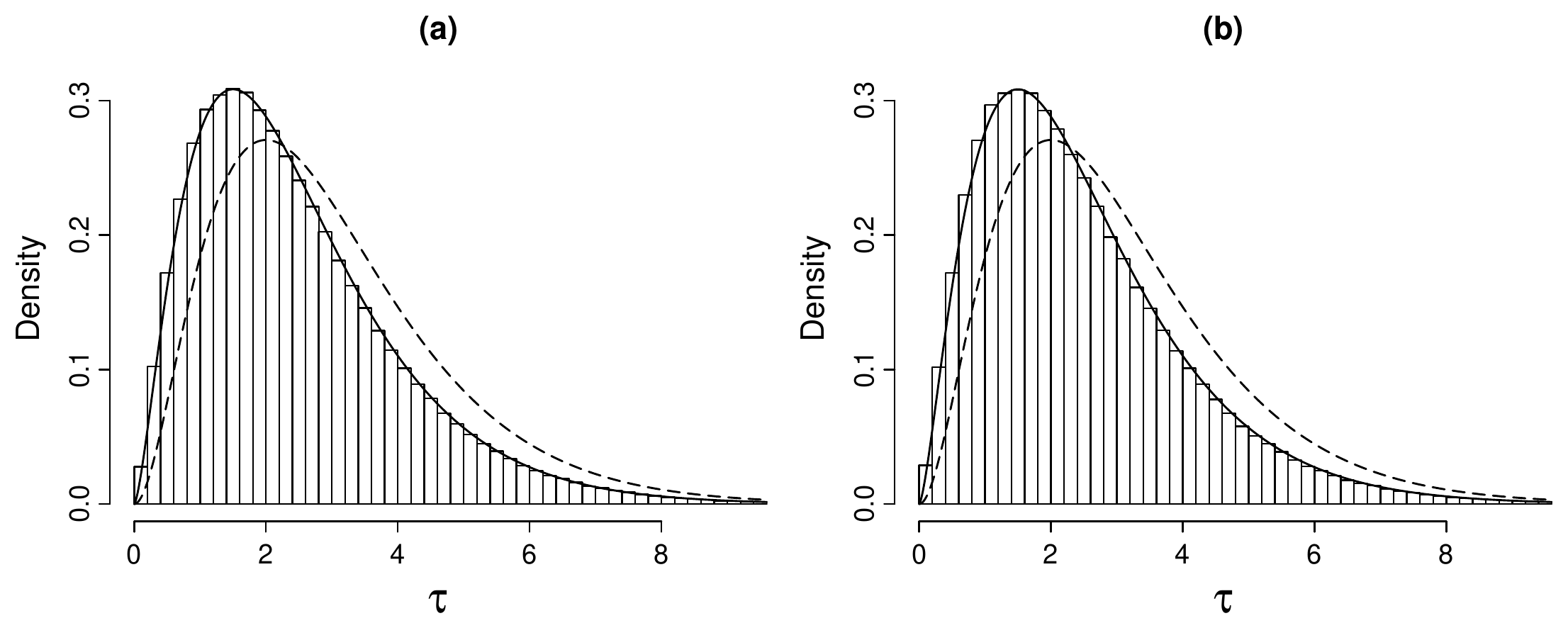}}
\caption{{\small Histograms representing samples from fiducial densities of a Poisson event rate}}
\end{center}
\end{figure}

We can see from Figures~3(a) and~3(b) that, although the posterior density for the event rate
$\tau$ is highly sensitive to which of the two prior densities for $\tau$ is used, there is almost
no difference in the fiducial density of $\tau$ depending on which of the two LPD functions of
$\tau$ is used.
Also, similar to what was the case for the two fiducial densities of $p$ in Figures~2(a) and~2(b),
the two fiducial densities of $\tau$ represented in these figures are both closely approximated by
the posterior density of $\tau$ that is based on the Jeffreys prior density for the problem of
interest.

Finally, we should place a minor caveat on this analysis that is similar to one that was placed on
the analysis of the example in the previous section, by observing that it may often be appropriate
to assess the variation in the fiducial density $f(\tau\,|\,x)$ over a wider range of LPD functions
of $\tau$ than is represented by the two LPD functions in equations~(\ref{equ13}) and~(\ref{equ14})
before any final conclusions about the precise nature of the post-data distribution of $\tau$ are
made.

\vspace{3ex}
\subsection{Inference about a multinomial distribution}

The version of Principle~2 outlined in Section~\ref{sec5}, i.e.\ the present version of this
principle, could be applied to the same problem of making post-data inferences about the parameters
of a multinomial distribution that was discussed in Section~6.3 of Bowater~(2019).
However, the results of this analysis would be very similar to the results that were obtained in
Bowater~(2019) when the older version of Principle~2 that was proposed in this earlier paper was
applied to the problem of inference in question. To be clear, it is not being asserted that the
results of these two analyses would be consistent with each other. Nevertheless, the differences
between the two sets of results being referred to would simply come down to how a sensitivity
analysis of the joint fiducial density of the parameters concerned over different LPD functions of
these parameters is affected by whether the density function $\omega_*(\theta_j\,|\,\gamma)$ that
is used in the definition of Principle~2 is, in effect, a conditional prior density or whether it
is the conditional posterior density that corresponds to this prior density. Therefore, for the
sake of wishing to avoid too much repetition of an earlier analysis of the problem of making
inferences about multinomial proportions that was presented in Bowater~(2019), a re-analysis of
this problem using the methodology of the current paper will not be presented here.

\vspace{3ex}
\section{What has been achieved?}

As well as clarifying to some extent how a fiducial statistic $Q(x)$ is defined when applying the
theory of organic fiducial inference to the most general type of inferential problems, this paper
has presented a modification to Principle~2 for defining the fiducial density
$f(\theta_j\,|\,\theta_{-j})$ relative to how this principle was specified in Bowater~(2019), and
it has been shown how this modification affects analyses of problems of inference that were
originally examined in this earlier paper.

\vspace{5ex}
\noindent
\textbf{References}

\begin{description}

\setlength{\itemsep}{1ex}

\item[] Bowater, R. J. (2018).\ Multivariate subjective fiducial inference.\ \emph{arXiv.org
(Cornell University), Statistics}, arXiv:1804.09804.

\item[] Bowater, R. J. (2019).\ Organic fiducial inference.\ \emph{arXiv.org (Cornell University),
Sta\-tis\-tics}, arXiv:1901.08589.

\item[] Bowater, R. J. (2020).\ Integrated organic inference (IOI):\ a reconciliation of
statistical paradigms.\ \emph{arXiv.org (Cornell University), Statistics}, arXiv:2002.07966.

\end{description}

\end{document}